\documentclass[11pt,preprint]{aastex}
\newcommand{\beq}{\begin{equation}}
\newcommand{\eeq}{\end{equation}}

\newcommand{\hi}{H{\sc i}~}
\newcommand{\hia}{H{\sc i}}
\newcommand{\kms}{km ${\rm s^{-1}}$~}
\newcommand{\kmsa}{km ${\rm s^{-1}}$}

\begin{document}

\title{The GALFA-H\hspace{0.15em}{\sc i} Survey: Techniques}
\author{J.~E.~G.~Peek\altaffilmark{1}, Carl Heiles\altaffilmark{1}}

%\noindent{J.~E.~G.~Peek, Carl Heiles}

\altaffiltext{1}{Department of Astronomy, University of California, Berkeley, CA 94720}

\begin{abstract}
We explain the entire process by which we conduct the Galactic Arecibo L-Band Feed Array \hi (GALFA-\hia) survey. The survey is a high resolution (3.4$^\prime$), large area (13000 deg$^{2}$), large Galactic velocity range ($-750$ to +750 \kmsa), high spectral resolution (0.18 \kmsa) survey of the Galaxy in the 21 cm line hyperfine transition of hydrogen conducted at Arecibo Observatory. We touch on some of the new Galactic science being conducted using the GALFA-\hi survey, ranging from High-Velocity Clouds to \hi narrow-line self-absorption. We explain the many technical challenges that confront such a survey, including baseline ripple, gain variation and asymmetrical beam shapes. To correct for these systematic effects we use various newly developed methods, which we describe in detail. We also explain the data reduction process step by step, starting with the raw time-ordered data and ending with fully calibrated maps. The effects of each step of the data reduction on the final data product is shown sequentially. We conclude with future directions for the ongoing survey.
\end{abstract}

\section{Introduction}
In 2004 the Arecibo L-Band Feed Array (ALFA) was installed on the Arecibo 305-meter radio telescope. While Arecibo's status as the largest single-dish antenna in the world had always made it a premier instrument in cm-wave radio astronomy, its ability to map large areas of sky had been hindered by the fact that it could only observe at one position at a time. One can consider the original L-band wide receiver, which covers a similar frequency range to ALFA, as a `single-pixel camera' with a single $3.4^\prime$ beam. ALFA, by comparison, monitors 7 positions at once, thus increasing the mapping speed of Arecibo by a factor of 7. This tremendous boost in efficiency led many different science teams to consider for the first time the possibility of mapping huge swaths of sky with Arecibo to create a new generation of high-fidelity, high-resolution sky atlases. These groups were divided by topic: pulsar (P-ALFA), extra-galactic (E-ALFA) and Galactic (GALFA) science. The GALFA consortium itself consists of three independent groups, or sub-consortia, divided by the radiative processes they measure: radio recombination lines (GALFA-RRL), continuum radiation (GALFACTS) and the 21 cm line transition of \hi (GALFA-\hia; \citealt{Goldsmith03}). 

Successful large \hi surveys of the Galaxy have been undertaken in the past, two of note being the Leiden-Argentina-Bonn (LAB) Survey \citep{Kalberla2005}, which is a completed full sky map of the Galaxy in \hia, and the International Galactic Plane Survey (I-GPS; \citealt{Taylor03}, \citealt{Stil06}, and \citealt{McC-G05}), an ongoing collaboration of interferometric observers compiling high resolution observations of the Galactic plane in \hi and other wavebands. GALFA-\hi is both large area (like the LAB) and high resolution (like the I-GPS), which, along with high spectral resolution and superb fidelity, allows it to contribute to a whole new range of Galactic science (see Table \ref{survcomp} and \S \ref{scigoal}).

In this paper we present a description of the GALFA-\hi survey in progress, including some of the science done with the survey (\S \ref{scigoal}), the overall signal chain (\S \ref{thesys}), the technical challenges posed by the survey (\S \ref{techchal}), the observing techniques we employed (\S \ref{obstech}), the way the data were processed (\S \ref{redproc}), and the future challenges for the survey (\S \ref{futdir}). For further discussion of the survey see \citealt{Stanimirovic2006}.

\begin{table}[htdp]
\begin{center}
\begin{tabular}{c c c c c c}
Survey & Area & Ang. Res.& Vel. Range & Vel. Res. & RMS\\
\hline
\hline
LAB & 41000 deg$^2$ & 36$^\prime$ & 850 \kms & 1.3 \kms & $\le$ 0.1 K \\
I-GPS & $\sim$1000 deg$^2$ & 1$^\prime$ -- 2$^\prime$ & 400 \kms & 1.3 \kms & 1 -- 3 K\\
GASS & $\sim$27000 deg$^2$ & 15$^\prime$ & 950 \kms & 0.8 \kms & 0.07 K \\
GALFA-\hi & 13000 deg$^2$ & 3.4$^\prime$ & 1500 \kms & 0.18 \kms & $\le$0.1 K \\
\hline
\end{tabular}
\end{center}
\caption{A comparison of a few modern, large-area Galactic \hi surveys. RMS is measured over a 1 \kms channel.}
\label{survcomp}
\end{table}
\section{GALFA-HI Science}\label{scigoal}

The GALFA-\hi survey was designed to study a broad range of topics under the umbrella of the neutral, Galactic Interstellar Medium (ISM). The GALFA white paper, \citet{Goldsmith03}, highlights 9 separate science projects that can be done with GALFA-\hia, 7 at high Galactic latitude and 2 and low Galactic latitude. Indeed, many science projects have been undertaken since the survey began, including observations of \hi narrow-line self-absorption in the Taurus \citep{Krco07}, studies of the disk-halo interface region \citep{Stanimirovic2006}, studies of the tip of the Magellanic Stream \citep{Stanimirovic08}, studies of high-latitude molecular clouds \citep{Gibson07}, the development of morphological distance measures to HVCs \citep{Peek07}, the discovery of low-velocity halo clouds (Peek et al. 2008), studies of interstellar turbulence \citep{Chepurnov06}, and the study of halo gas around M33 \citep{Putman07}. Further science work with the GALFA-\hi survey has begun on a wide variety of topics including the search for \hia-rich Milky Way dwarfs, the investigation of mysterious Galactic hyper-pressure objects and the study of magnetized \hi filaments in spiral arms. 

\section{The System}\label{thesys}

\subsection{The AO 305m}
\begin{figure}
\begin{center}
\includegraphics[scale=.65, angle=0]{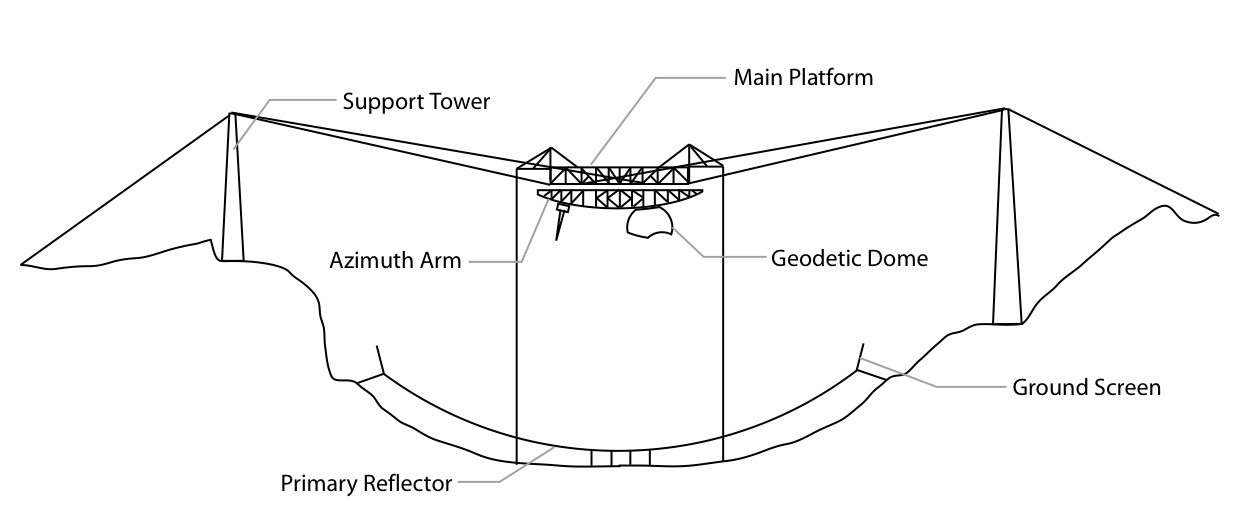}
\caption{A diagram of the Arecibo telescope.\label{aodiag}}
\end{center}
\end{figure}

The GALFA-\hi survey is conducted on the Arecibo 305-meter telescope (see Figure \ref{aodiag}), located at the Arecibo Observatory (AO), South of Arecibo, Puerto Rico and is part of the National Astronomy and Ionosphere Center (NAIC). The primary reflector is fixed spherical cap of 305 m diameter and 265 m radius of curvature. Suspended above the primary reflector is a triangular platform that carries a circular track upon which the azimuth arm rotates. The azimuth arm carries a geodetic dome (also called the Gregorian dome), which houses secondary and tertiary optics to correct for spherical aberration. The entire system can track objects from declination -1$^\circ$ to 39$^\circ$ declination with $\sim 5^{\prime\prime}$ accuracy. ALFA is located on a rotating turret inside the dome at the focal plane of the telescope. The details of the 305m are discussed in detail in the Arecibo Technical and Operations Memos Series (ATOMS) and on the NAIC website, \texttt{http://www.naic.edu}.

\subsection{ALFA}
\begin{figure}
\begin{center}
\includegraphics[scale=.70, angle=0]{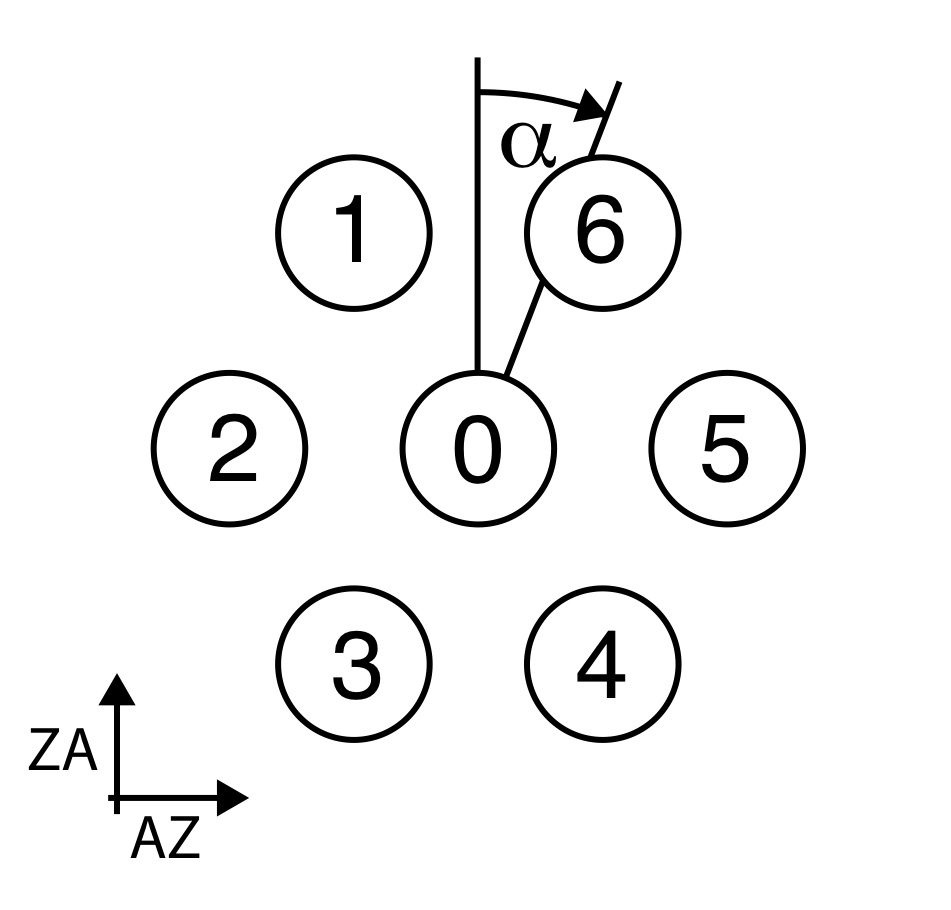}
\caption{The layout of the 7 ALFA beams in azimuth -- zenith angle coordinates with the definition of the rotation angle, $\alpha$.\label{beamlayout}}
\end{center}
\end{figure}

ALFA is a 7-beam feed array designed to allow Arecibo to perform surveys at a greatly increased rate. The receiver frontend was built under contract to the Australia Telescope National Facility of the Australian's Commonwealth Scientific and Industrial Research Organization. The beams are arranged on the sky in a hexagon with a beam 0 in the middle (see Figure \ref{beamlayout}). Each beam is slightly elliptical along the zenith-angle direction, approximately $3.3^\prime \times 3.8^\prime$ full-width half-maximum (FWHM) on the sky, with a gain of $\sim$11 Jy/K for beam 0 and $\sim$8.5 Jy/K for beams 1-6. Each beam is divided into independent dual-linear polarizations, for a total of 14 independent signals. ALFA can be tuned from 1225 to 1525 MHz, which includes the hyperfine transition of \hi at 1420.405 MHz. For more design and implementation details consult the ALFA technical memo series, \texttt{http://www.naic.edu/alfa/memos/}.

\subsection{GALSPECT}
The GALFA-\hi spectrometer, GALSPECT, consists of 14 intermediate frequency (IF) to quadrature baseband downconverters, 7 dual polarization spectrometer boards, a CompactPCI computer Board, an Agilent 8648A synthesizer, all powered by an uninterruptible power supply. The system is designed to take 14 IF signals split and amplified from the ALFA signal chain and generate 14 32-bit spectra each second over 7679 channels covering $\sim$ 7 MHz as well as 14 32-bit spectra each second over 512 channels covering $\sim$ 100 MHz (see {\texttt http://seti.berkeley.edu/galfa} for details). The wide-band spectra are used to calibrate the narrow-band `science' spectra (see \S \ref{bfsfs}). The spectrometer has superb channel-to-channel isolation due to the use of a fourier-transform polyphase filterbank and loses less than 1\% of data from `dead time' during data readout. The spectra are tagged with system parameters (telescope pointing, IF parameters, etc.) through the AO Shared Common Random Access Memory Network (SCRAMNet) and are written each second over the AO network to a large server for data storage and reduction.

\section{Technical Challenges}\label{techchal}
	\subsection{Baseline Ripple}\label{baseripch}
\begin{figure}
\begin{center}
\includegraphics[scale=.65, angle=0]{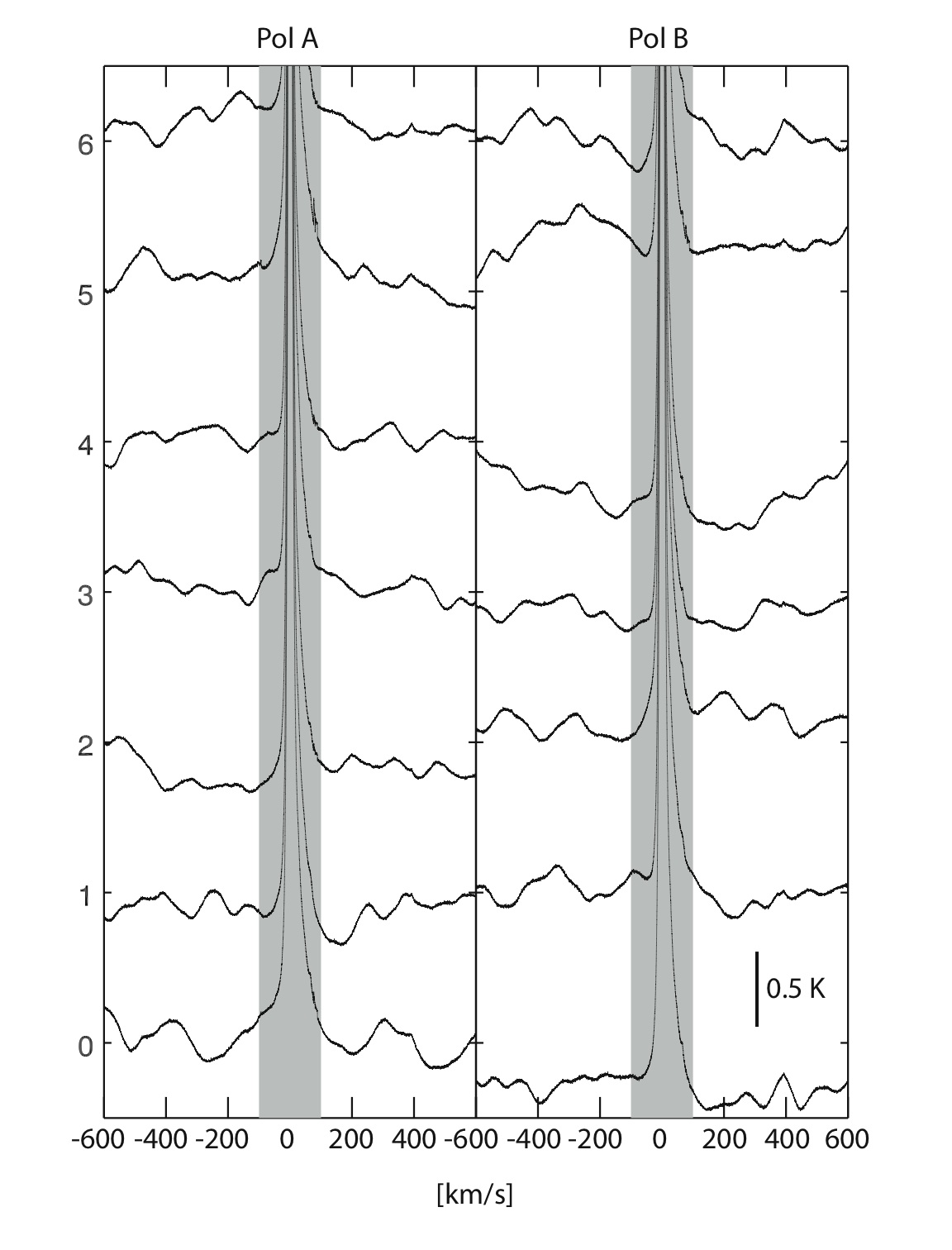}
\caption{An example of the baseline ripple in each of the 14 ALFA receivers from the integrated spectra over 6 hours at fixed azimuth, zenith angle and ALFA rotation angle. The gray area represents the domain in which the HI contributes noticeably to the average spectrum.\label{exripple}}
\end{center}
\end{figure}

It was known since long before the GALFA-\hi project was undertaken that spectra taken at the Arecibo 305m (and, indeed, almost all radio telescopes) are typically contaminated by what is called a `baseline ripple' (see, for example, \citealt{Padman74} and references therein). This ripple, also called fixed-patten noise, is typically a low-level wave that permeates the entire spectrum, with an amplitude of $\sim 0.2$K (see Figure \ref{exripple}). This contamination to the spectra is caused by reflections inside the Arecibo 305m superstructure and geodetic dome. Any signal that takes more than a single path to the receivers will cause a spurious increase in the amplitude of the auto-correlation function (ACF) in the channel corresponding to the delay between the paths. In the case of a signal that takes both a standard path to the receivers and one that has an extra reflection between the geodetic dome and the main reflector, for instance, the path length difference is on the order of 300 meters, which corresponds to a delay of 1 $\mu$s. This 1 $\mu$s delay causes an increased ACF at the 1 $\mu$s lag which, in turn, generates a ripple with a period across the spectrum of 1 MHz. Such a ripple has $\sim 3$ periods across the $\sim 1000$ \kms that the Galactic \hi occupies, and thus is capable of significantly corrupting our final data product if not addressed adequately.

\subsection{IF Bandpass}\label{ifbp}

In radio heterodyne spectroscopy, the measured spectrum is the product of the radio-frequency (RF) power and the IF gain spectra; to obtain the RF power spectrum, one must divide the on-source measured spectrum (the `on spectrum') by the IF gain spectrum. This is usually accomplished by dividing by a reference spectrum (the `off spectrum'), which is obtained by moving off in frequency or position. The 21 cm line is ubiquitous on the sky, so moving to a displaced position is ineffective. Moving off in frequency combines the baseline ripples (see \S \ref{baseripch}) at the two frequencies, which further contaminates the data. 

\subsection{Gain Variation}
\begin{figure}
\begin{center}
\includegraphics[scale=.50, angle=0]{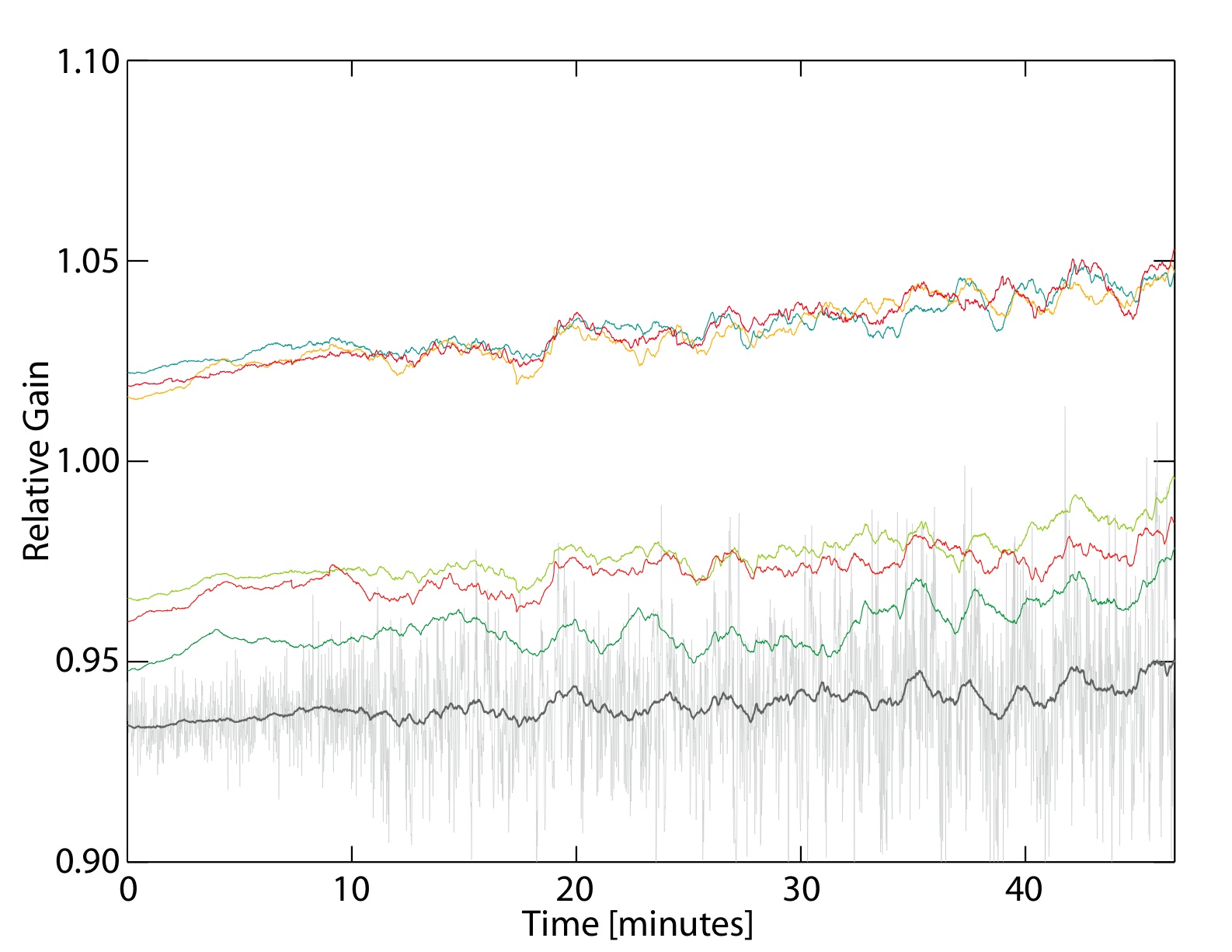}
\caption{The relative gains of two consecutive days of observation, taken at identical positions. Each coincident integration is compared for each polarization-averaged beam, plotted in 7 different colors. We find the relative beam gain between the two days at each second by fitting to the \hi spectrum taken on the first day plotted against the \hi spectrum taken on the second day, where the relative gain is the slope of the line. Each relative beam gain is averaged over 100 seconds to smooth the gain trends. The smoothed beam 0 gain data is plotted in dark gray, with the unsmoothed beam 0 relative gains in light gray for comparison. The noise in the unsmoothed gain fits changes over the course of the observation time as the observations leave the Galactic disk and the \hi amplitude decreases significantly.\label{gainex}}
\end{center}
\end{figure}

To make a clean map from data taken with a single dish radio telescope, the overall gains need to be accurately calibrated. These gains can vary significantly from day to day, based on both the performance of the telescope and the receiver (see Figure \ref{gainex}). These gains may even vary during long observations. It is possible to calibrate these variable gains with the calibration noise diodes in the ALFA system, but there is no guarantee that these diodes themselves generate a fixed amplitude at the receiver. Uncalibrated gains lead to striation in the maps and inaccurate \hi amplitudes. 

\subsection{Beam Shape}\label{beamshape}
\begin{figure}
\begin{center}
\includegraphics[scale=.75, angle=0]{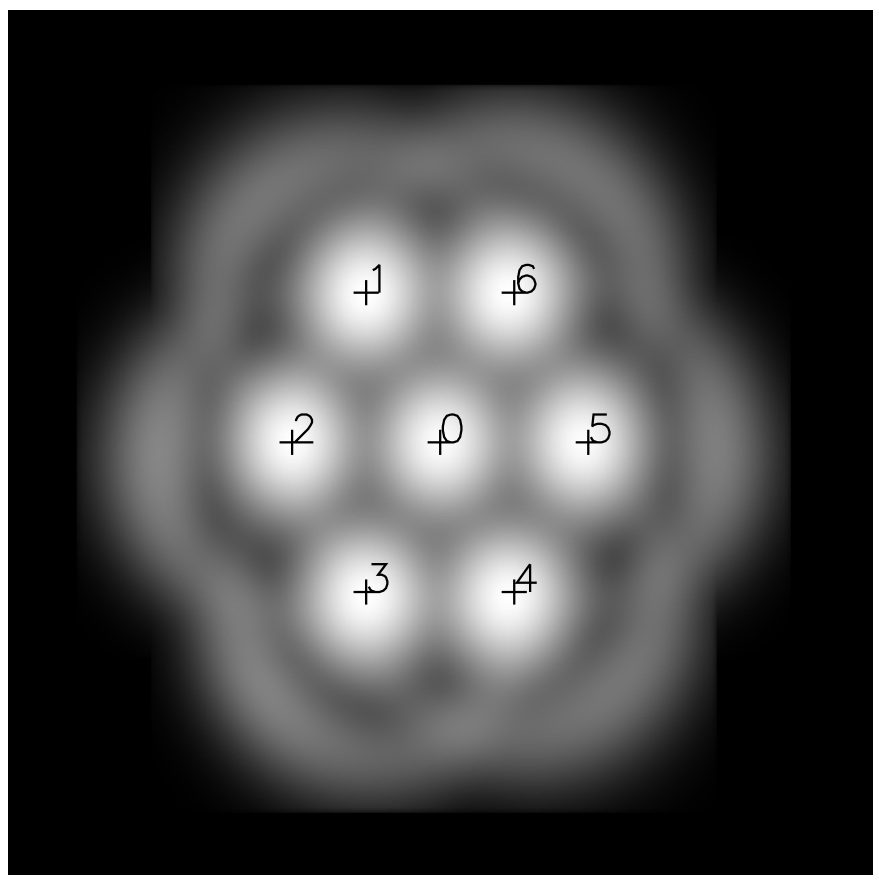}
\caption{In image of the response function of each of the 7 ALFA beams. Note the significant sidelobe response in beams 1 -- 6. \label{slplot}}
\end{center}
\end{figure}

Each ALFA beam has a significantly different beam shape, because of the very fast optics of the telescope. In particular, each beam has significant first and second sidelobes which depend upon the relative displacement of the beam to beam 0 (see Figure \ref{slplot}). The first sidelobes are $\sim 5^\prime$ offset from the main beams with overall efficiencies, $\eta_{SL}$, near 20\% for beams 1 to 6. Since we do not intend to Nyquist sample the sky with each beam, but rather with a combination of all the beams, we will end up with a map that has an ill-defined associated beam pattern. This means that many of the methods devised for removing the effect of beam shape from maps (Maximum Entropy, CLEAN) cannot be used and we must devise our own method for removing the effects of these sidelobes.

	\subsection{Radio Frequency Interference}
While Arecibo is located in a relatively remote part of the world and is in the Radio Astronomy Coordination Zone in Puerto Rico, both of which reduce the amount of radio frequency interference (RFI), observers at AO do frequently encounter RFI. The GALFA-\hi experiment is relatively narrow band, as compared to extragalactic or pulsar studies, which also reduces the exposure to RFI problems. Nonetheless, RFI removal is important to have well-calibrated data. 

\section{Observing Techniques}\label{obstech}
A suite of observing techniques are used in the GALFA-\hi survey, which we address here in turn. Many of the single-beam techniques these modes were based upon are described in \citet{ONeil02} and \citet{Stanimirovic02b}.

\subsection{Observing Philosophy}
When the GALFA-\hi project was started, the importance of both a cohesive final data product and science-oriented observing approach were recognized. These two goals could easily come into conflict if not treated carefully: too much freedom for observers to pursue individual projects could yield data that would be difficult to combine into a final survey product, too many constraints on the observations and scientific goals would be hard to pursue. 
To facilitate the creation of a user accessible final survey, all of the GALFA-\hi data are taken in a single data-taking mode, with a fixed bandwidth, resolution and frequency setting. This is made possible by the GALSPECT correlator, which simultaneously has narrow enough channels (0.18 \kms) for the most stringent Galactic \hi science criteria (\hi narrow-line self-absorption, e.g., \citealt{Krco07}) as well as enough bandwidth (+750 to -750 \kms) to observe the fastest moving Galactic HI objects (very high velocity clouds, e.g., \citealt{Peek07}). In this way all projects can be served by the same spectrometer setting. It is also important to allow individual observers to image higher-priority projects first. To observe the totality of the Arecibo sky, even at the shallowest possible level, would require more than 1700 hours of observing, thus delaying the publication of particular science of interest significantly. The decision to allow individual PIs to to choose science targets in this ``jigsaw puzzle'' strategy has indeed facilitated the timely release of data and the inclusion of students in the survey process.

\subsection{Least-Squares Frequency Switching}\label{lsfs}

Least-squares frequency switching (LSFS) is fundamentally different from frequency switching. In frequency
switching, one takes two spectra---and on-line and an off-line
spectrum---and (1) takes the difference, to remove the continuum
(frequency-independent) system temperature, which is large; and (2)
divides the difference by the off-line spectrum, which corrects for the
frequency-dependent system gain. This technique introduces two
artifacts. First, the system temperature is not independent of
frequency, and in particular at Arecibo it has variations (``wiggles'')
on the $\sim 1$ MHz scale. Thus, when taking the difference in (1), the
wiggles in both the on-line and the off-line spectra appear in the
difference spectrum; these are uncorrelated, so the wiggle amplitude
increases by $\sim \sqrt{2}$. Second, the system gain depends mainly on
IF frequency, and the division in (2) reliably corrects for this
IF gain component. However, the system gain also varies with RF
frequency to some extent, and---as for the wiggles---RF gain
variations in the off-line spectrum appear in the final result. LSFS
avoids some of these problems because (1) it avoids the need for
subtracting an off-line spectrum, and (2) it derives the IF gain
alone, so that the final derived spectral gain errors come only from
those in the RF spectrum.

In LSFS, one sets the local oscillator (LO) frequency
to a number of different values so that the RF power spectrum and the IF
gain spectrum can be evaluated as distinct entities using a
least-squares technique \citep{Heiles07}. In practice, at the beginning
of each observation we spend 10 minutes recording data at a single sky
position at a number of different frequencies. We used the
minimum-redundancy schema with 7 LO settings. In this ``MR7'' schema,
the LO frequency separations range from 1 to 39 times the the minimum
separation $\Delta f$. One needs the maximum separation to be a
reasonable fraction of the total spectrometer bandwidth so that the
full-bandwidth gain spectrum is accurately defined. With our large
number of channels, this means that $\Delta f$ must be much larger than
the channel spacing $\delta f$. We chose ${\Delta f \over \delta f} =
225$, which gave a maximum separation of 90\% of the full spectrometer
bandwidth.

When ${\Delta f \over \delta f} > 1$, one needs a reliable way to
degrade the resolution of the original 7679-channel spectra to reduce
the number of channels in the LSFS technique to a manageable
number. One applies the LSFS technique to these degraded-resolution
spectra to derive a degraded-resolution IF gain spectrum. Finally, one
interpolates the degraded-resolution gain spectrum to regain the
original number of spectral channels. We used standard Fourier series to
both degrade the original resolution and to recover the original
resolution, reducing the number of channels in the degraded-resolution
spectra to 512.

\subsection{Scanning Methods}

	\subsubsection{Drift Scanning}
\begin{figure}
\begin{center}
\includegraphics[scale=.50, angle=0]{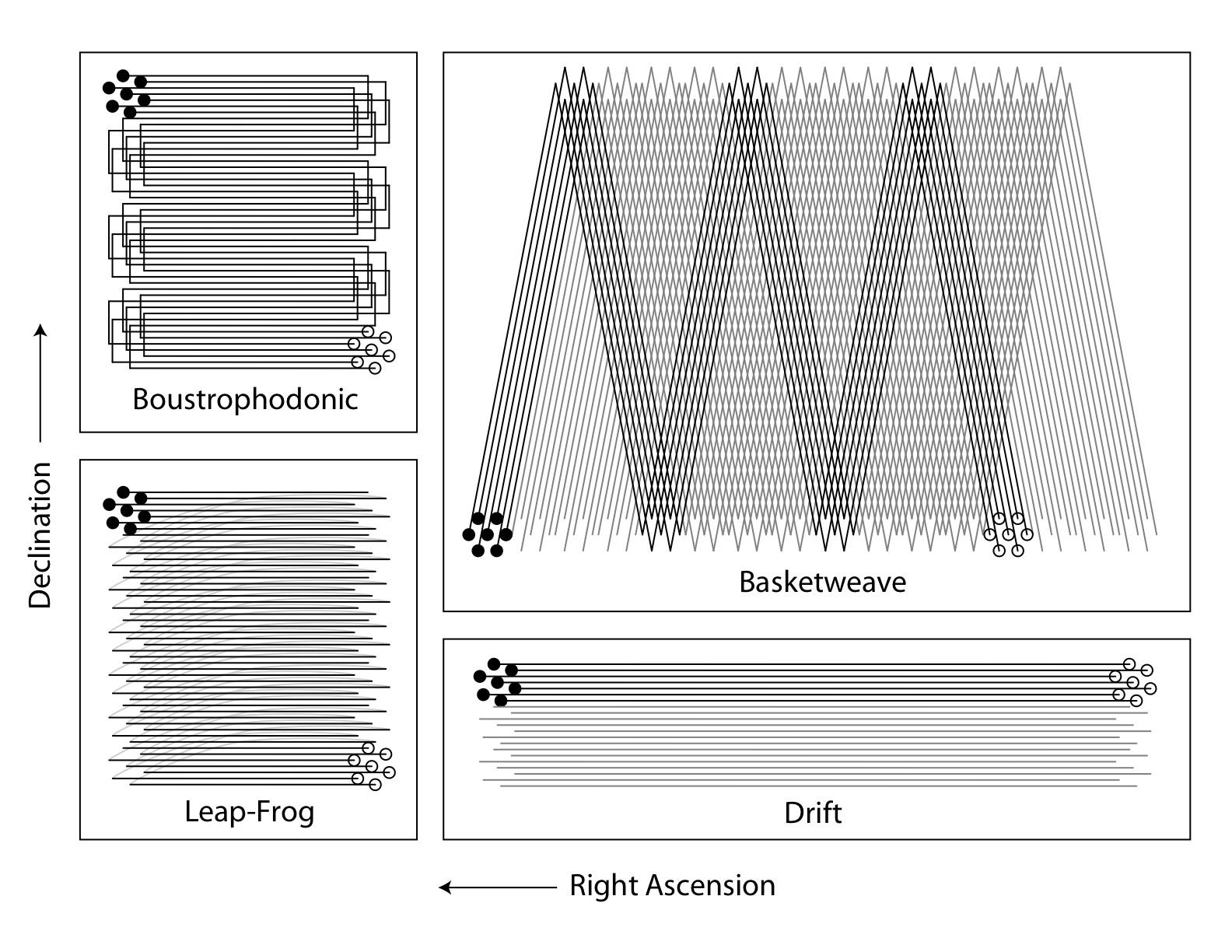}
\caption{The four modes of scanning used by GALFA-\hia. In each example, the empty circles represent the initial position of ALFA for a single day's observation and the filled circles represent the final position of ALFA for the day's observation. In the drift and basketweave diagrams gray lines represent complementary scans taken on other days of observations. In the leap-frog diagram the gray lines represent the return slew of the telescope to the beginning of the right-ascension range to start another fixed-azimuth scan.\label{scanpattern}}
\end{center}
\end{figure}

The simplest method of observing is the drift-scanning method. Indeed, this method traces back to the dawn of radio astronomy as the telescope is simply left at a fixed position at transit (in our case at azimuth = 180 or azimuth = 0) and the sky is allowed to drift by. To space the beams equally, the optimal angle for the ALFA beams is $\sim 19^\circ$, as shown in Figure \ref{scanpattern}. This scanning pattern has the advantage that the maximum number of telescope parameters are held fixed, which minimizes variability in baseline ripple. This pattern also has a number of disadvantages. Firstly, it is unwieldy for regions that are relatively large in declination (or small in right ascension). Secondly, since the scans are all parallel on the sky, there is no easy way to cross-calibrate the scans, as they do not intersect (see \S \ref{bw}). Thirdly, the telescope has limitations on the zenith angle at which it can point, so regions near zenith are inaccessible in the standard mode. It is possible to drift scan away from transit to get declination range near zenith, but one sacrifices constancy of baseline ripple pattern. This method is used primarily for the Arecibo Legacy Fast ALFA Survey (ALFALFA; \citealt{Giovanelli05}), which is a wide-area survey with which GALFA-\hi commesally observes.

	\subsubsection{Leapfrog Scanning}	
Leapfrog scanning is a variation on drift scanning, in which the limitation to large regions is overcome by stepping to different fixed azimuths over the course of an observation (see Figure \ref{scanpattern}). Essentially, the sky is allowed to drift by at a fixed azimuth over some range in right ascension. Then, the telescope slews forward to the beginning of the right ascension range at a different declination and a new fixed azimuth scan is taken, and so forth. In this way each individual scan across a region has the same fixed parameters as in a drift scan, but smaller regions can be more easily observed in longer contiguous blocks. This method allows for much deeper scanning of a region. This method is primarily used by the Arecibo Galaxy Environments Survey (AGES; \citealt{Auld06}), with which GALFA-\hi commesally observes.
	\subsubsection{Boustrophedonic Scanning}
The word `Boustrophedonic' comes from the Greek for `ox-turning', which describes scanning the sky as an ox would plow a field. This method is also called RA/Dec scanning (see \citealt{ONeil02}). The telescope slews in increasing right ascension, steps in declination, slews in decreasing right ascension, steps in declination and so forth (see Figure \ref{scanpattern}). Parallactic angle tracking must be used to maintain parallel, equally spaced beams throughout this process. This allows the observer to examine regions with arbitrary aspect ratio at any place on the Arecibo sky and at any integration time per beam. It also is not limited to declinations far from zenith. It has the disadvantages that telescope pointing parameters are not held constant and that intersection points are not generated. This method has been used to map small regions by individual GALFA-\hi PIs.
	\subsubsection{Basketweave Scanning}\label{bw}
The preferred mode for GALFA-\hi observations is basketweave or `meridian-nodding' scanning. In this mode the telescope is kept at the meridian and is driven up and down in zenith angle over the chosen declination range. Each day the starting position of these nods is displaced in right ascension such that the entire region is covered, as in Figure \ref{scanpattern}. There are six distinct drive speeds (or `gears') which can generate equally spaced tracks with each of the seven beams: three with an ALFA rotation angle of $0^\circ$ and three with an ALFA rotation angle of $30^\circ$. This observing method has the advantage that it generates a large number of intersection points among beam tracks at which the relative gain of the beams can be measured (as in \citealt{Haslam81}). This allows us to greatly reduce the striation in the map caused by gain variation. It does this while maintaining a fixed azimuth, which significantly limits the fluctuations in baseline ripple. 

A somewhat modified version of the basketweave can be employed at declinations near zenith where the traditional basketweave scan cannot be used.  In this method, called `azimuth wagging', the same scan pattern on the sky is achieved, but by positioning the telescope at near 90 or 270 degrees azimuth. This method is inferior to the standard basketweave in that the azimuth, zenith angle and ALFA rotation angle must all change continuously during the observation.

\section{The Reduction Process}\label{redproc}

\subsection{Bandpass Fitting with SFS}\label{bfsfs}
As discussed in \S \ref{ifbp}, we need to correct each spectrum for the IF gain.  At the beginning of each observing period, we use LSFS to obtain this IF gain spectrum and we use this spectrum for the whole period's data. This works because the gain spectrum is stable and changes only imperceptibly. In fact, the IF gain spectrum doesn't change from day to day as long as the IF electronics chain is exactly identical. We can ensure this invariance by setting all aspects of the IF electronics, including all attenuator and amplifier gain values, to standard values.

	\subsection{Initial Calibrations}\label{initcal}
\begin{figure}
\begin{center}
\includegraphics[scale=.50, angle=0]{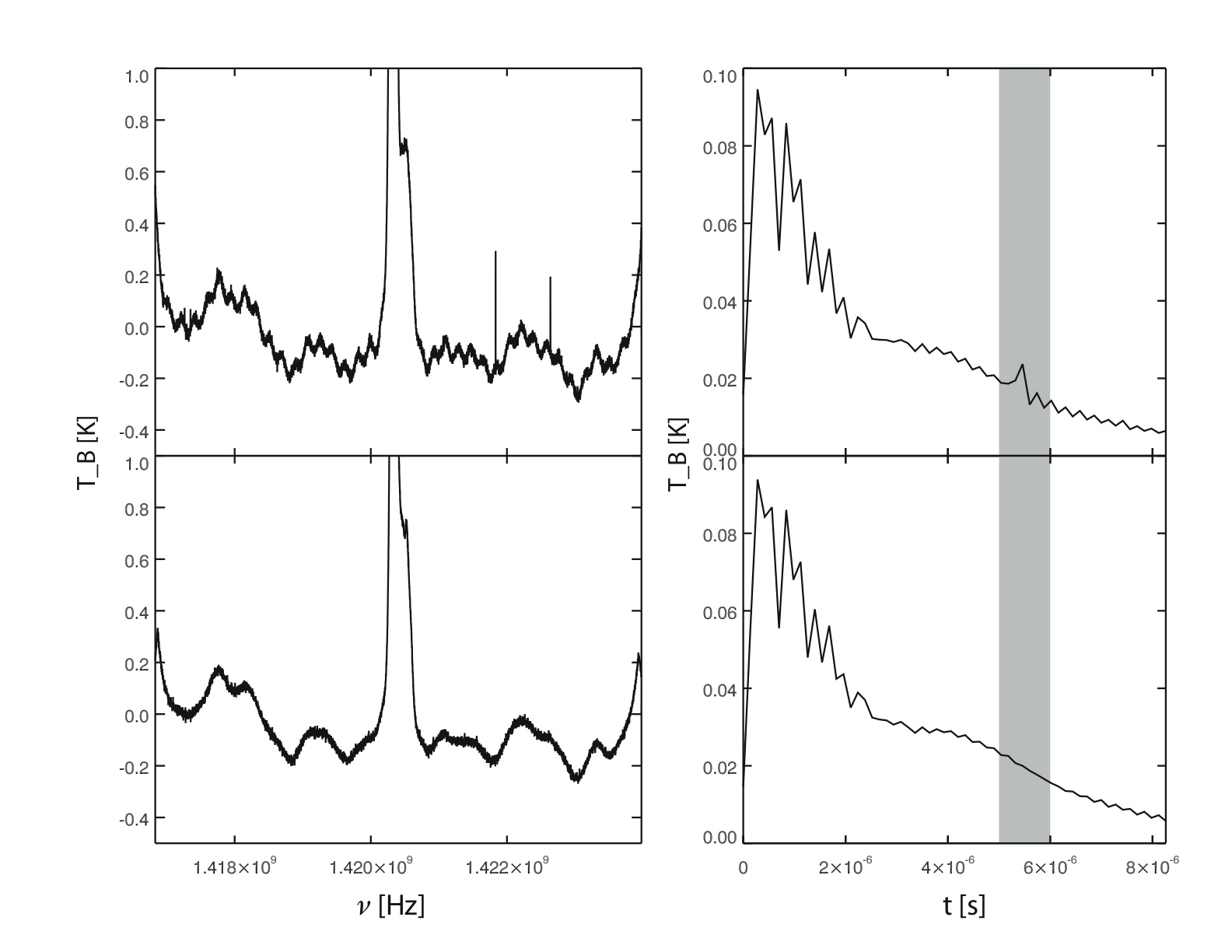}
\caption{The plots on the left are time averaged spectra from a drift scan observations in the original state (top) and with the ripple removed (bottom). The plots to the right are the Fourier transform of the spectra, with the region where the ripple was removed highlighted.\label{refl_ex}}
\end{center}
\end{figure}

After the effects of the bandpass are removed, the data are searched for possible contamination by reflections in the signal chain (see Figure \ref{refl_ex}). Any impedance mismatch between cables or signal transmitting hardware can cause echoes within the system. Similar to the reflections in the superstructure and geodetic dome of the telescope, these reflections in the cables produce excess power at lags equivalent to the reflection's differential travel time, and therefore cause ripples in the baseline of the spectrum. These reflections have the very nice quality that they are each only a single Fourier mode: as the reflection geometry is only one dimensional, there is no room for complex reflection patterns as there is in the telescope superstructure. This ripple is therefore very easily removed by examining the data in Fourier space and simply removing the clear reflection tone. After this cleaning is done the data can be shifted into the local standard of rest (LSR) frame.

	\subsection{Ripple Subtraction}\label{ripsub}
Removing the ripple caused by the reflections in the telescope superstructure and geodetic dome is much more difficult than removing the effects of reflections within the signal chain. This is because the ripple is not dominated by a single Fourier component, but rather by a large range of Fourier modes, with associated delay ranging from 0.5 to 1 $\mu$s. This ripple was examined in some detail in \citet{heiles05}. This work showed that the ripples are 100\% polarized, uncorrelated from receiver to receiver, relatively static over time and dependent upon ALFA rotation angle, telescope azimuth and zenith angle. 

\begin{figure}
\begin{center}
\includegraphics[scale=.55, angle=0]{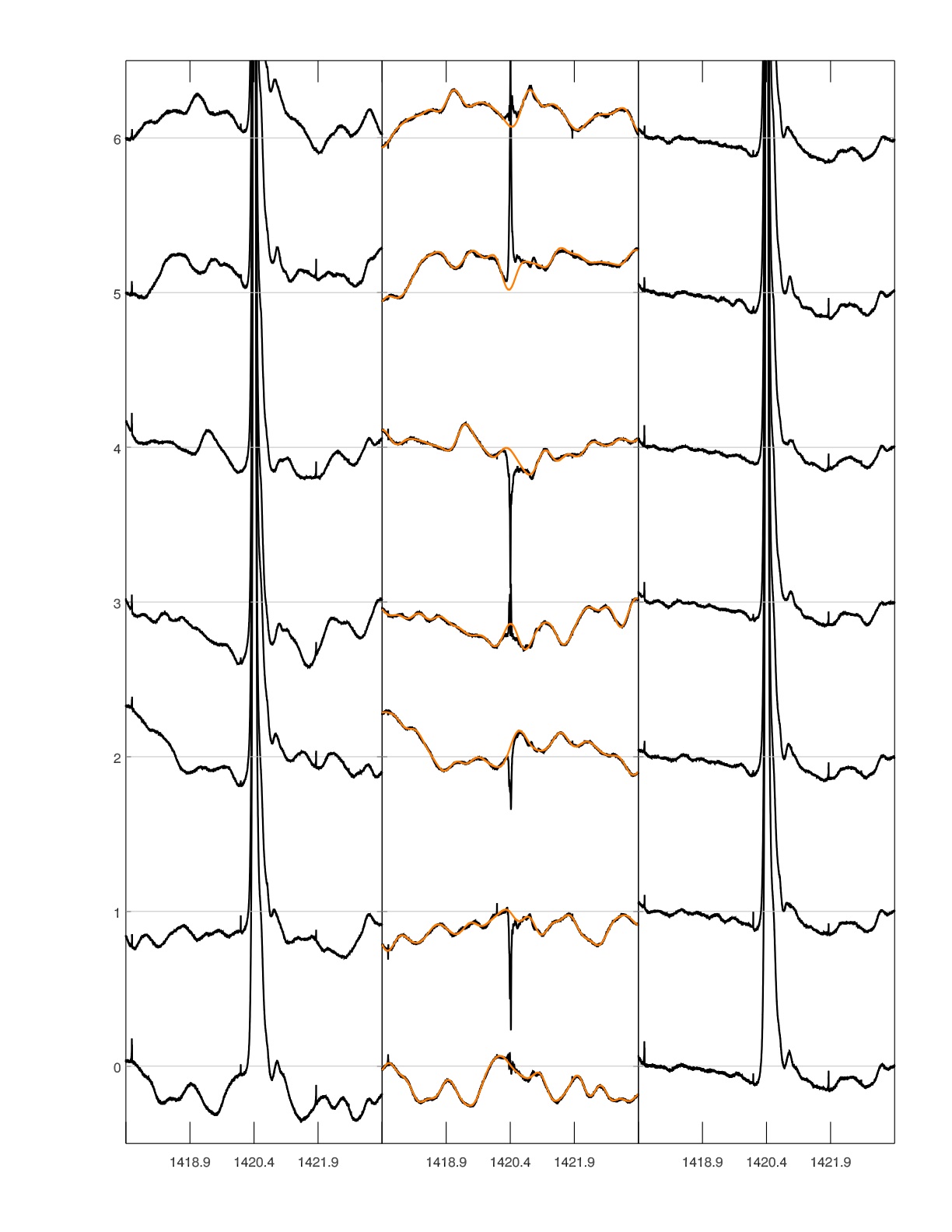}
\caption{An example of the fixed-pattern noise. The left panel shown the uncorrected fixed-pattern noise in each of the seven beams, averaged over polarization and over a few hours of observation. A gray line at zero is shown for comparison. The middle panel shows these same spectra less their average. Plotted over this is the fit to these spectra, as in the ripple component of Equation \ref{ripple}. The right panel shows the original spectra less the fit in the center panel. The ripple has been significantly reduced.\label{fpnex}}
\end{center}
\end{figure}

To reduce the effects of the ripple we use the fact that each beam has an uncorrelated and relatively fixed ripple over the course of an observation in our two main observing modes, drift scanning and basketweave scanning. In the case of drift scanning all telescope position parameters are held fixed, so we expect no variation in the ripple. In the case of basketweave scanning only the zenith angle is changed, which causes an average change of $\sim 25 \%$ in the ripple over the full range of zenith angle, which we ignore. After averaging the two polarizations within a beam and applying a simple overall gain calibration, we reduce the effect of these ripples by first generating the average spectrum, $T_{avg}\left(\nu \right)$ for all 7 beams over the course of a single day's observation. We then fit the difference between $T_{avg}\left(\nu \right)$, and the average over a single day's observation, $T_{n}\left(\nu \right)$, with 
\begin{eqnarray}\label{ripple}
T_{n}\left(\nu \right) - T_{avg}\left(\nu \right) &=& \sum_{i=1}^m \left( a_{i}sin\left(t_i\nu\right) + b_icos\left(t_i\nu\right)\right)  \\
&& + \sum_{j=1}^l \left(\left(\frac{\partial T}{\partial\alpha}\right)_{\nu_j}\Delta\alpha_nd\left(\nu - \nu_j\right) + \left(\frac{\partial T}{\partial\delta}\right)_{\nu_j}\Delta\delta_nd\left(\nu - \nu_j\right)\right) \nonumber
\end{eqnarray}
where, $t_i$ ranges over the observed baseline ripple frequencies (0.5-2 $\mu$s) and $\nu_j$ ranges over the frequencies where disk \hi is typically observed. $d$ is equivalent to the Dirac delta function, to avoid confusion with declination, $\delta$. The first component of the fit is the baseline ripple, the second component is the residual \hi contribution. By fitting the residual \hi contribution with this Taylor expansion in $\alpha$ and $\delta$ we are able to accurately capture the \hi residuals while only fitting 2 parameters per channel, rather than the 6 required if we were to fit each beam individually. By subtracting only the baseline ripple component of the fit for each beam, we are able to reduce the amplitude of the baseline ripple in each beam by $\sim \sqrt{7}$ without disturbing the \hi data. An example of this method is shown in Figure \ref{fpnex}.

	\subsection{Intersection-Point Calibration}\label{intpnt}
To determine the relative gains of each beam over each day of observation we use the information that at places where beam tracks intersect they must observe the same region of sky, and therefore any differences in their spectra must be dominated by variation in the gain parameters. This method is most effective in basketweave scans where there are a very large number of intersection points - a region 2 hours across in right ascension and 17 degrees in declination has almost $10^5$ intersection points. If we wish to determine the relative gains in a set of drift scans, we must conduct other observations that intersect the drift scans, as drift scans are parallel on the sky.

Once the intersection points are found, we determine the relative gains at these points by comparing the amplitudes of the observed \hi spectrum by each beam. Each beam is assumed to have an overall gain for each day's observation plus some variability with time during the observation we parameterize as a sum of Fourier components. Since only relative gains are measured, the overall gain solution must average to unity. We construct a least-squares fit to the measured relative gains at intersection points for the various day and time dependent gain parameters and then apply these parameters to the entire data set to remove the relative gains.
	\subsection{Gridding}\label{grid}
Once the spectra, or time-ordered data (TOD), have been fully calibrated, we apply them to a grid in the image plane. This grid can theoretically be in any projection in the World Coordinate System as described by \citet{CG2002}, but we typically project to Cartesian coordinates in right ascension and declination with orthogonal graticules. We use a Gaussian sampling kernal with FWHM = $3^\prime$ to place flux on the grid, which smoothes out any pixels that may have been missed by the observing pattern. Our method is anchored by the IDL code SDGRID.PRO, a code significantly adapted by Timothy Robishaw from work by the authors, which was in turn based upon the IDL gridzilla implementation by Sne\v zana Stanimirovi\'c. SDGRID.PRO uses the core logic described in \citet{Dickey02} and many of the methods developed for Parkes gridzilla \citep{Barnes01} and the AIPS procedure SDGRID.  After the gridding is complete a final amplitude correction is applied to make the observations consistent with the lower resolution Leiden-Dwingaloo Survey (LDS).
	\subsection{First Sidelobe Calibration}\label{fsc}
As we mentioned in \S \ref{beamshape} a significant fraction of the signal observed by ALFA comes in through the first sidelobe, particularly in the off-axis beams 1-6. This will cause significant asymmetrical beam smearing. To counteract this effect we use a strategy similar to that employed to correct the LDS, as outlined in \citet{HKBM96}. Because we have seven different beams, rather than a single convolving beam, we must modify their method somewhat. In essence, we `re-observe' at each of the locations of the TOD by assuming the sky is identical to our data cube from \S \ref{grid}. In this re-observation we use only the measured sidelobe response function at each of these TOD, effectively building a `contamination' spectrum for each position. These contamination spectra are then subtracted from the original spectra,
\beq
T^\prime\left(\nu, t\right) = T\left(\nu, t\right)/\eta_{\rm MB} - \eta_{\rm SL}T_{C}\left(\nu,t\right)/\eta_{\rm MB},
\eeq
Where $T\left(\nu\right)$ are the original TOD, $T_{C}\left(\nu,t\right)$ are the `contamination spectra', and $\eta_{\rm MB}$ and $\eta_{\rm SL}$ are the main beam and side-lobe efficiencies, respectively. These corrected spectra, $T^\prime\left(\nu, t\right)$ are then re-gridded to make a final map. While there are second-order effects that stem from using the contaminated map as the true sky image, these effects are relatively small, of order $\eta_{\rm SL}^2$.

\section{Data Processing: An Example Region}\label{exr}
As an example of the GALFA-\hi survey observations and data processing methods we have described in the previous sections, we show here a specific region observed over several months in 2006. The entire region is $20^\circ \times 17^\circ$, centered on 17h +$8^\circ$, and was chosen by the investigators (Mary Putman and Sne\v zana Stanimirovi\'c) because HVC complex C is visible in the region. It was observed with the basketweave method (i.e.~\S \ref{bw}) twice, to achieve a higher signal-to-noise. The data were reduced with the method outlined above.

\begin{figure}
\begin{center}
\includegraphics[scale=.70, angle=0]{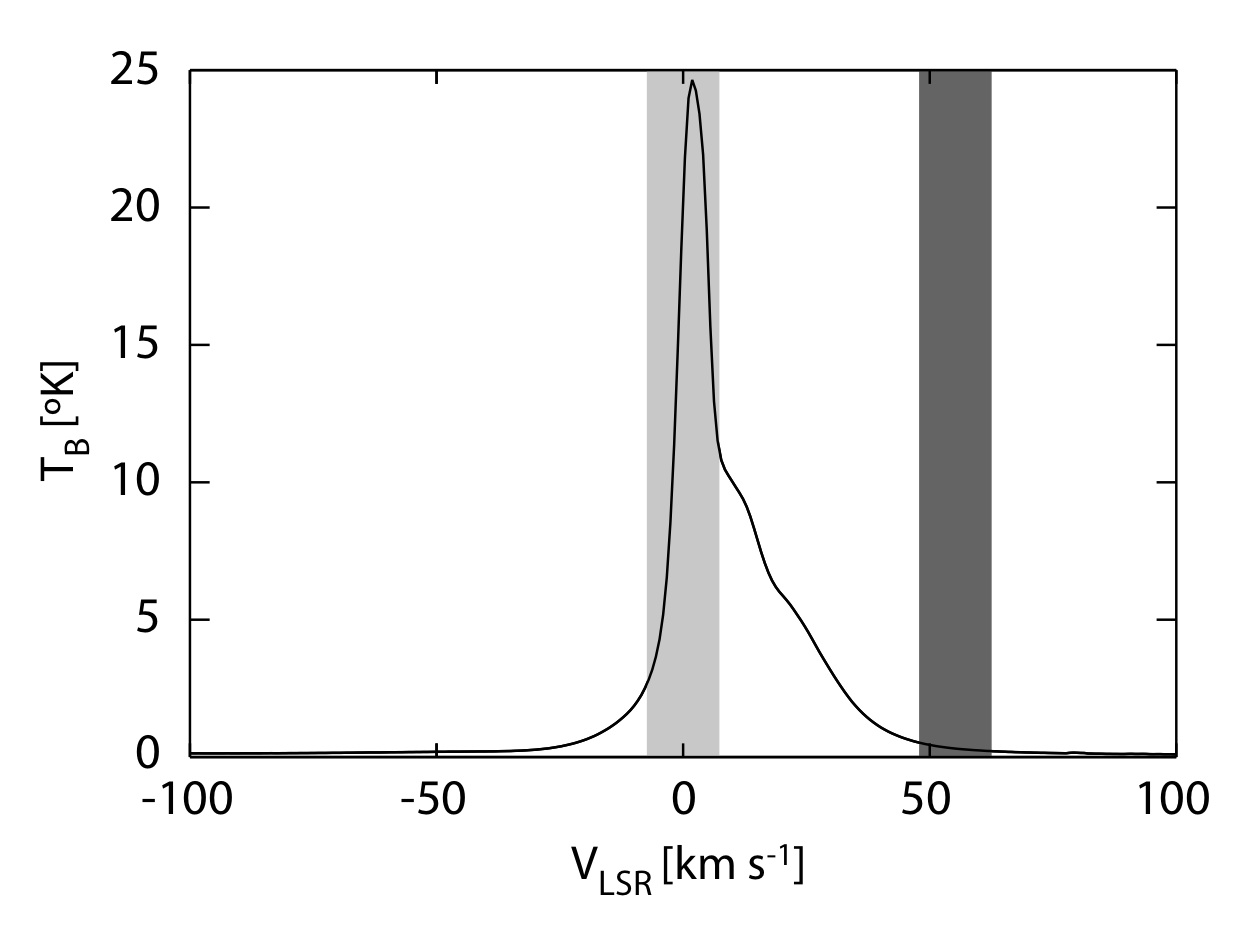}
\caption{The average spectrum over the region discussed in \S \ref{exr}. The dark gray area represents the velocities integrated over in the off-line map, while the light gray area represents the velocities for on-line map. \label{ex_spec}}
\end{center}
\end{figure}

\begin{figure}
\begin{center}
\includegraphics[scale=.68, angle=0]{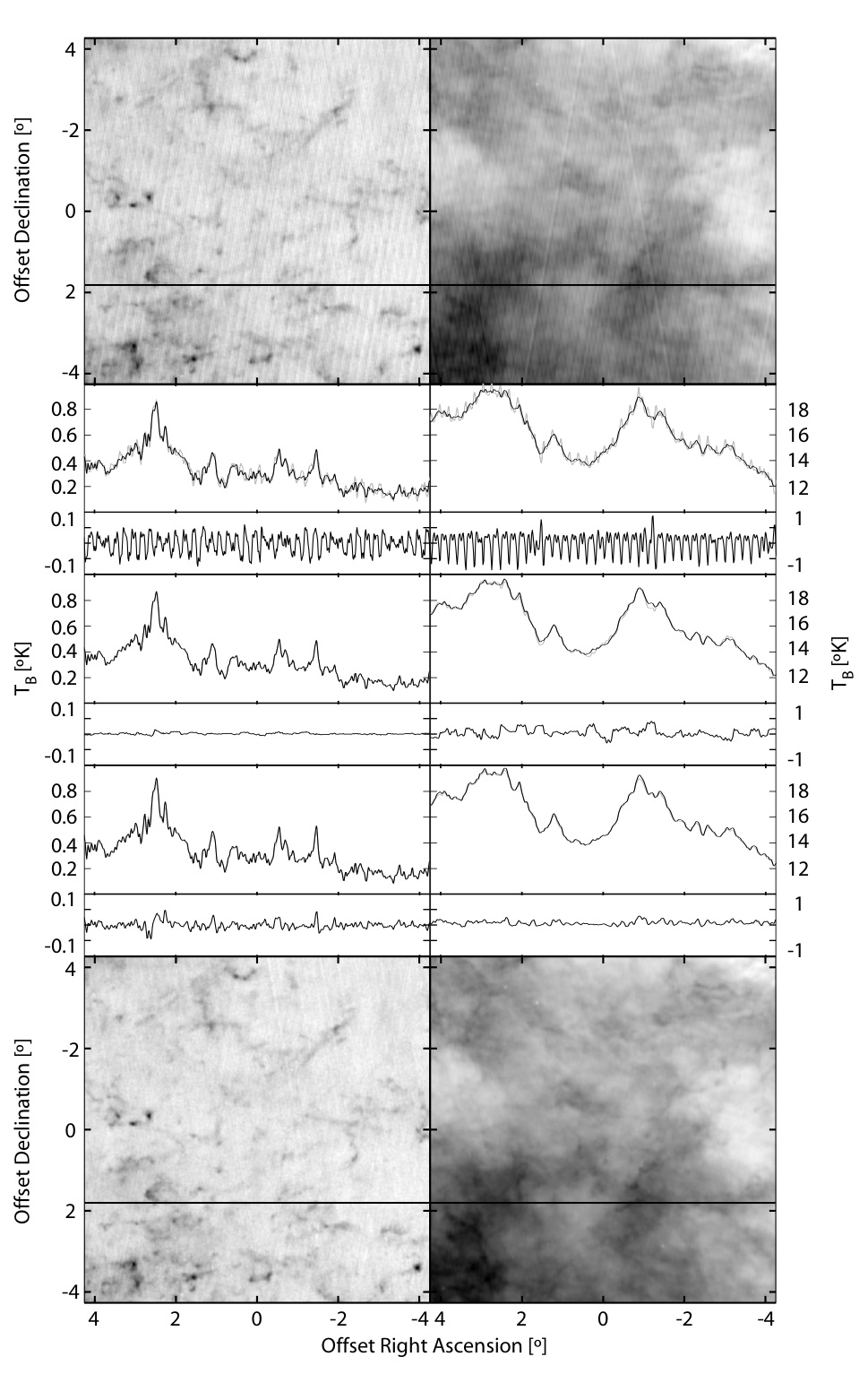}
\caption{An example region of the sky towards 17h 52m +$10^\circ 21^\prime$. The left and right columns represent the dark and light regions in Figure \ref{ex_spec}, respectively. See \S \ref{exr} for details. \label{ex_reg}}

\end{center}
\end{figure}

In this example we only investigate a single survey cube, which is $512 \times 512~ 1^\prime$ square pixels and in the standard Cartesian projection in equatorial coordinates. This cube is centered on 17h 52m +$10^\circ~ 21^\prime$. Figure \ref{ex_spec} shows the average spectrum over the cube and the two velocity ranges we consider. We isolate these velocity ranges to demonstrate the effect of each stage of the data reduction on both strong \hi (on-line) and weak \hi (off-line) maps. We demonstrate the effects of each stage of reduction in Figure \ref{ex_reg}, by showing a map of the uncalibrated data at the top, fully-calibrated data at the bottom, and slices of the data through each step of calibration in between.

The top two panels of Figure \ref{ex_reg} show the integrated intensity map of the off-line and on-line regions without any calibration past the initial calibrations described in \S \ref{initcal}. The strong striation along the scan patterns from poorly calibrated gains and baseline ripple are quite evident. The following three pairs of plots (and corresponding residual plots) below show the effect of applying the ripple subtraction (\S \ref{ripsub}), intersection-point calibration (\S \ref{intpnt}), and the first-sidelobe calibration (\S \ref{fsc}), respectively. Each plot is of a slice at constant declination (indicated in the intensity maps by a solid line) of the maps at the various stages of reduction. In the first plot the a slice of the uncalibrated map is shown in gray, with the data calibrated for baseline ripple shown in black and the difference between the two plotted below. The effect is quite noticeable in both plots. In the case of the off-line plot, the cleaning is primarily due to the ripple reduction itself, as the amplitude of the features on the sky is similar to the amplitude of the ripple. In the on-line slice the cleaning is mostly due to the rough gain calibration applied at this stage. The second pair of map slices shows the ripple-cleaned data in gray and the intersection-point calibrations in black, with the difference between the two plotted below. The off-line data show a very modest change due to this calibration, whereas the much brighter on-line data show significant changes in some regions. The third pair of map slices show the data calibrated for ripple and intersection points in gray and the data fully calibrated in black, with the difference between the two plotted below. This final step of first-sidelobe calibration has most effect at the peaks of the slices where data had been erroneously scattered, and has a noticeable effect on both off-line and on-line slices. The last pair of maps shows the same region and velocity range as the first set of maps, but with all three data reduction methods applied. The off-line data are significantly improved, but the effects of baseline ripple are still quite evident. Given the low dynamic range of the map ($\sim 1$K), it is not surprising that residual ripple on the order of 0.1 K (see Figure \ref{fpnex}) is still visible. The on-line data are also drastically improved, and even with very close inspection it is very hard to determine the presence of systematic artifacts. This is not to say that systematic effects are not present, but they are certainly greatly reduced by these methods.

\section{Future Directions}\label{futdir}
We have made significant strides towards reliable, artifact-free data, and we expect that these data will be a useful resource for the astronomy community. To further this goal we intend to continue to pursue known sources of systematic error in our data, including the effects of more distant sidelobes and RFI.

\subsection{Further Sidelobe Calibration}
While the primary source of stray radiation is undoubtably the first sidelobes, the ALFA beam pattern is known to have significant power in second and higher-order sidelobes. These sidelobes roughly follow a Airy pattern model with elliptical cross section \citep{C-M2002}. Second sidelobes are more distant from the center of the observation, so while the contamination amplitude will be limited, the signal in the second sidelobe will be typically be more divergent from the main beam than in the first sidelobe. In cases where there is very fast variation in the sky signal this second sidelobe can have a significant effect, and we believe that it is currently the dominant contribution to uncorrected sidelobe radiation. 

To correct for the second and higher-order sidelobes these sidelobes must be carefully measured. Some raster maps have been made of the second sidelobe response, but they are significantly incomplete. To parameterize the basic components of second sidelobe we will measure it with the `spider scan' method outlined in \citet{Heiles04}. We will then add the second sidelobe response to our calibration method described in \S \ref{fsc}.
	\subsection{Ground Screen Stray Radiation Calibration}
The Arecibo primary reflector is surrounded by a 16 m ground screen (see Figure \ref{aodiag}), to prevent thermal radiation from the ground from entering the main beam of the telescope when it points at high zenith angle. As our dominant observing modes are on the meridian, the ground screen is only illuminated towards the edges of our declination range, near $-1^\circ$ and $38^\circ$. The ground screen does not follow the shape of the spherical cap of the primary reflector, so while it does prevent significant ground signal from entering the main beam, it also significantly distorts the main beam, adding distant sidelobes. If these sidelobes were adequately modeled through geometric optics, we could use methodologies similar to those described in \S \ref{fsc} to reduce the effect of the ground screen. 
	\subsection{Automated RFI Rejection}
At present, RFI rejection is not part of the standard data reduction process. Typically, there is very little noticeable RFI in an image, but what RFI exists occupies a single channel within the science spectrum. Once the data are processed and corrected to the LSR frame, these spikes can be `smeared out' in frequency space, thus further contaminating the data and making RFI rejection more difficult. Therefore to implement RFI rejection in the future we intend to scan each incoming science spectrum with a median filter to look for significant outliers. To preserve the integrity of the spectra, only these outliers would be subject to excision. We expect that this process will remove most RFI found in our data sets. 
\section{Conclusion}

In this work we described the systems (\S \ref{thesys}), observing modes (\S \ref{obstech}) and and data reduction methods (\S \ref{redproc}) used to produce the GALFA-\hi survey. We demonstrated the efficacy of these techniques (\S \ref{exr}) and discussed methods for improving our final data product in the future (\S \ref{futdir}). We found that we were capable of drastically reducing the impact of many of the systematic effects on the data cubes, and that we are well on our way to an optimally reduced final data product. 

\section{Acknowledgements}

The authors would like to thank the many, many people who made this work, and the ongoing survey it represents, possible. We thank Paul Goldsmith, Tom Bania and John Dickey for early science conceptual work on the GALFA project. The GALFA-\hi survey and TOGS team, Mary Putman, Snezana Stanimirovi\v c, Steven Gibson, Eric Korpela, Kevin Douglas, Jana Grcevich and Min-Young Lee were crucial in conceiving of, proposing and conducting the survey. We thank Marko Krco for his contribution of the GALFA-\hi archiver. We thank Tim Robishaw for his work on SDGRID.PRO. We thank the incredibly helpful and dogged Arecibo staff, including the telescope operators who worked with us at great length to develop our observing protocols. In particular we would like to thank Phil Perillat and Mikael Lerner for their contributions to the GALFA-\hi data reduction pipeline and observing codes, respectively. We thank Arun Venkataraman for assistance in data handling and storage. We would also like to thank Dan Wertheimer and his team for designing and building the GALFA spectrometer, GALSPECT. We would like to especially thank the late Jeff Mock, without whose dedication and brilliance this project could not have happened. This work was supported in part by NSF grant AST-0406987 and NSF Collaborative Research grant AST-0707679,070758,0709347.

\bibliographystyle{apj}

%\bibliography{/Users/goldston/Documents/publications/mybib}
%\begin{thebibliography}
%\end{thebibliography}

\end{document}